\documentclass [12pt]{article}
\usepackage {graphicx}
\usepackage[american]{babel}

\begin{document}

\title{The simplest map with three-frequency quasi-periodicity and quasi-periodic bifurcations}

\author{A.P. Kuznetsov and Yu.V. Sedova}

\maketitle
\begin{center}
\textit{Kotel'nikov's Institute of Radio-Engineering and Electronics of RAS, Saratov Branch,\\
Zelenaya 38, Saratov, 410019, Russian Federation}
\end{center}

\begin{abstract}
We propose a new three-dimensional map that demonstrates the two- and
three-frequency quasi-periodicity. For this map all basic quasi-periodic
bifurcations are possible. The study was realized by using method of Lyapunov
charts completed by plots of Lyapunov exponents, phase portraits and
bifurcation trees illustrating the quasi-periodic bifurcations. The features
of the three-parameter structure associated with quasi-periodic Hopf
bifurcation are discussed. The comparison with non-autonomous model is carried out.
\end{abstract}

\textit{PACS:} 05.10.-a, 05.45.-a

\section{Introduction}

Quasi-periodic oscillations are widespread in nature and technology
\cite{Ble88,Lan96,Pik01,An07}. Examples of quasi-periodic behavior are found in electronics \cite{Gla88},
neurodynamics \cite{Izh00}, astrophysics \cite{Lam85}, physics of lasers \cite{Mel91}, geology \cite{Did11} and
other areas of science.

In the simplest case the quasi-periodic oscillations are characterized by
the presence of two incommensurable frequencies. In the phase space
attractors corresponding to such oscillations have the form of invariant
tori \cite{Ani95,Shi01,Bro10}. There are more complicated cases of greater number of
incommensurable frequencies - then multi-frequency oscillations are observed
which correspond to invariant tori of higher dimension. Invariant tori
may undergo a variety of bifurcations. In this case we speak about
quasi-periodic bifurcations \cite{Bro08,Bro08a,Vit11}. The basic ones are the next three
bifurcations:
\begin{itemize}
  \item the saddle-node bifurcation. Collision of stable and unstable tori leads
to the abrupt birth of higher dimensional torus.
  \item the quasi-periodic Hopf bifurcation. Torus of higher dimension is born
softly.
  \item the torus-doubling bifurcation.
\end{itemize}
Currently there are no reliable identification algorithms for such
bifurcations and their search is executed by means of the characteristic
dependence of Lyapunov exponents on parameter \cite{Bro08,Bro08a,Vit11}.

While examining the quasi-periodic bifurcations the selection of the model
for studying plays a crucial role. Some aspects can be studied based on
non-autonomous systems. However, such systems form separate, special
class. As concerning the known autonomous models, the number of low-dimensional variants is not very large. Several
examples of suitable generators are proposed recently. The first example,
obviously, is a Chua's circuit, which is described by specific piecewise
linear characteristic \cite{Mat87}. In \cite{Nis06} a generator based on the modified
Bonhoffer - van der Pol system is offered. Also we know a modification of
climate Lorenz model -- the Lorenz-84 low-order atmospheric circulation
model \cite{Shi95}. Another radiophysical
example implemented as a four-dimensional system  is a modified Anishchenko-Astakhov generator \cite{Ani05,Ani06,Ani07}. In
\cite{Kuz10,Kuz13,Kuz15,Kuznet15} a family of simple generators of quasi-periodic oscillations
described by three-dimensional models was suggested and studied (including experiment).
There have been studied cases of autonomous dynamics as well as generator
under external forcing and dynamics of coupled oscillators \cite{Ani06,Ani07,Sta15,Kuznet13}.

It is known that flow systems (differential equations) are quite difficult
for research, especially it concerns the study of so delicate effects like
quasi-periodic bifurcations. Therefore it is natural to deal with a simpler
model such as discrete maps. In this case the image of two-frequency
oscillations in the phase space is an invariant curve. As for flows the same
object emerges in Poincar\'{e} section of torus. (Thus frequently such invariant curves are also referred to as tori.) The higher is the map
dimension the higher may be dimension of the torus. Most
recently there have been undertaken appropriate investigations of model
maps. In paper \cite{Kuz12} the four-dimensional system of two coupled universal
maps with Neimark-Sacker bifurcation is examined. In \cite{Sek14,Kam14} there was
studied four-dimensional map representing two logistic maps with delay. The
authors of the papers \cite{Hid15,Hid15a} undertook analogous study of six-dimensional
system in the form of three logistic maps with delay. In \cite{Adi13} the
six-dimensional model which represents two coupled discrete versions of R\"{o}ssler
system is examined.

However, the dimension of above-mentioned systems is too large compared to the dimension which is formally necessary to observe two- and three-frequency quasi-periodicity. An exception
is the paper \cite{Dem14} that study the three-dimensional system in the form of
three coupled in a ring logistic maps with a specific coupling. Nevertheless, analysis of the problem in this paper seems insufficient. The noteworthy model from \cite{Bro08,Bro08a,Vit11} was used for studying the quasi-periodic bifurcations, but its
construction is quite complex formal procedure. Although the model describes
the very important points, the discussed picture is still not quite
complete. (The authors carried out only two-parameter analysis, at that they
focused mainly on the description of resonance 1:5.)

It may be marked also the paper \cite{Ani94} where the coupled logistic maps with
quasi-periodic forcing are studied. The research concerns bifurcations and
mechanisms of transition to chaos through the destruction of
three-dimensional torus. However, not all the essential points have been
investigated in detail. Moreover, this model belongs to the class of
non-autonomous systems, i.e. systems with external forcing.

Thus, there is a problem to construct autonomous models in the form of maps
with the minimum necessary dimension equal to three that allows to study the properties
of quasi periodic dynamics and quasi periodic bifurcations.
In the present paper such model is represented by a discrete analogue of
quasi-periodic generator \cite{Kuz13,Kuz15}. We undertake research of the structure of the parameter space focusing on the phenomena associated with quasi-periodic Hopf
bifurcation.

\section{Torus map construction}

Let us consider the generator of quasiperiodic oscillations \cite{Kuz13,Kuz15}:
\begin{equation}
\label{eq1}
\begin{array}{l}
 \ddot {x} - (\lambda + z + x^2 - \beta x^4)\dot {x} + \omega _0^2 x = 0, \\
 \dot {z} = b(\varepsilon - z) - k\dot {x}^2. \\
 \end{array}
\end{equation}
The multiplier ahead of the derivative $\dot {x}$ contains the parameter
$\lambda $ which characterizes the depth of a positive feedback in the
oscillator, the nonlinear term $x^2$ stimulating the excitation of oscillations
and the term $x^4$ responsible for the saturation of the oscillations at
large amplitudes. The dynamics of the variable z can occur linearly at a
speed $b$ or undergo the nonlinear saturation due to the term $k\dot {x}^2.$
\begin{figure}[!ht]
\centerline{
\includegraphics[height=6cm, keepaspectratio]{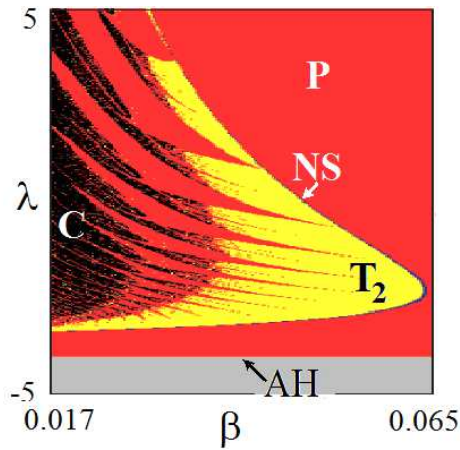}}
\caption{Lyapunov chart for generator of quasi-periodic oscillations (\ref{eq1}), $b =
1$, $\varepsilon = 4$, $k = 0.02$, $\omega _0 = 2\pi $.}
\end{figure}

The model (\ref{eq1}) has an equilibrium $x_0 = y_0 = 0,\,\,z_0 = \varepsilon .$ It
is easily verified that this equilibrium undergoes Andronov-Hopf bifurcation
AH at $\lambda = - \varepsilon $ leading to the birth of a limit cycle. With
increasing of parameter $\lambda $ the Neimark-Sacker bifurcation NS
consisting in the birth of a two-dimensional torus is possible. In Fig.1 we
demonstrate the Lyapunov chart for system (\ref{eq1}) where different colors
indicate the areas of periodic regimes $P$, the two-frequency
quasi-periodicity $T_2 $ and chaos $C$. We can see the curve of Neimark-Sacker
bifurcation $NS$ with the adjoined set of Arnold tongues immersed in the
region of quasi-periodic oscillations.

Let us rewrite equations (\ref{eq1}) in the form of first-order system
\begin{equation}
\label{eq2}
\begin{array}{l}
 \dot {x} = y, \\
 \dot {y} = (\lambda + z + x^2 - \beta x^4)y - \omega _0^2 x, \\
 \dot {z} = b\left( {\varepsilon - z} \right) - ky^2, \\
 \end{array}
\end{equation}
and construct a discrete analog of equations (\ref{eq2}). For this purpose we use a
substitution for time derivatives by finite differences similarly to
\cite{Zas88,Zas07,Arr93}. The transition to finite differences will provide some additional
characteristic time scale -- a discretization step, which usually leads to
new types of dynamics. The result is a model that we call a torus map:
\begin{equation}
\label{eq3}
\begin{array}{l}
 x_{n + 1} = x_n + h \cdot y_{n + 1} , \\
 y_{n + 1} = y_n + h \cdot \left( {(\lambda + z_n + x_n^2 - \beta x_n^4 )y_n
- \omega _0^2 x_n } \right), \\
 z_{n + 1} = z_n + h \cdot \left( {b(\varepsilon - z_n ) - ky_n^2 } \right).
\\
 \end{array}
\end{equation}
Here $h$ is discrete time step. Note that for the first equation we use
semi-explicit Euler scheme, i.e. we take the value of the variable $y$ in
$\left( {n + 1} \right)$-th moment. This discretization usually leads to
more physically correct models \cite{Mor05}.

\section{Properties of torus map with quasi-periodic dynamics}

Let us study the obtained map. We use the same set of parameters as for
Fig.1 and will gradually increase the discretization parameter $h$. In the
center of Fig.2 a numerically calculated Lyapunov chart for system (\ref{eq3}) at
the value $h = 0.05$ is shown. Different colors in the chart denote the
following regions defined in accordance with the spectrum of Lyapunov
exponents $\Lambda _1 ,\,\Lambda { }_2,\;\Lambda _3 $:

\noindent
a) $P$ -- periodic regimes (cycles), $\Lambda _1 < 0,\,\Lambda { }_2 <
0,\;\Lambda _3 < 0$;

\noindent
b) $T_2 $ -- two-frequency quasi-periodicity, $\Lambda _1 = 0,\;\Lambda { }_2
< 0,\;\Lambda _3 < 0$;

\noindent
c) $T_3 $ -- three-frequency quasi-periodicity, $\Lambda _1 = 0,\;\Lambda {
}_2 = 0,\;\Lambda _3 < 0$;

\noindent
d) $C$ -- chaotic regimes, $\Lambda _1 > 0,\;\Lambda { }_2 < 0,\;\Lambda _3
< 0$;

\noindent
e) $HC$ -- hyperchaotic regimes, $\Lambda _1 > \;\Lambda { }_2 > 0,\;\Lambda
_3 < 0$;

\noindent
f) $D$ -- a divergence of trajectories.

\begin{figure}[!ht]
\centerline{
\includegraphics[height=10cm, keepaspectratio]{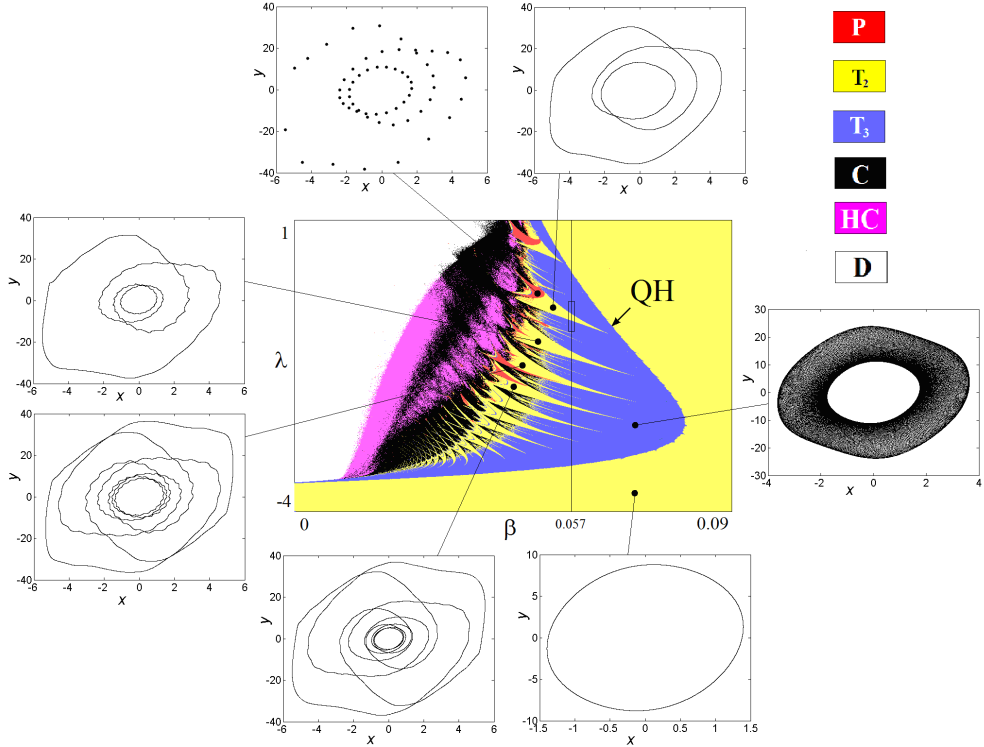}}
\caption{Lyapunov chart of the torus map (\ref{eq3}) and typical phase portraits. $QH$
is a line of quasi-periodic Hopf bifurcation. Discretization parameter value $h =
0.05$.}
\end{figure}

Due to the smallness of the discretization parameter $h$ the structure of
the chart is partly qualitatively similar to that of the original flow system (\ref{eq1}).
(The smaller $h$ the better an approximation to the flow system.) However
the discretization leads to replacement of periodic regimes of flow system (\ref{eq1}) by two-frequency regimes and two-frequency - by three-frequency ones in the Fig.1. Thus
in Fig.2 we can see a picture of tongues of two-frequency regimes that in
configuration are similar to traditional Arnold tongues. The mentioned set of
tongues immerses in a region of three-frequency tori. The tongues in Fig.2 correspond to resonant two-frequency tori lying on the surface of
three-frequency torus.

In Fig.2 we show examples of phase portraits in the various points of the
parameter plane. Below and right to the line QH torus looks like a simple
oval. In the three-frequency region this oval is smeared. Inside the tongues
of two-frequency regimes the attractors have the form of closed invariant
curves. In different tongues on the plain $\left( {x,y} \right)$ these
curves differ in the number of turns around the origin. Such curves are
replacing the simple limit cycles in the Poincar\'{e} section of the original model
(\ref{eq1}) and their shape indicates that discretization even with a small step $h$
leads to a specific modification of the observed structure.

In turn inside tongues of two-frequency tori the areas of periodic regimes
arise. In this case on the invariant curve of complex shape there is a set
of points of appropriate long-period cycle, i.e. such regimes are resonant
with respect to the corresponding two-frequency torus.

\begin{figure}[!ht]
\centerline{
\includegraphics[height=9cm, keepaspectratio]{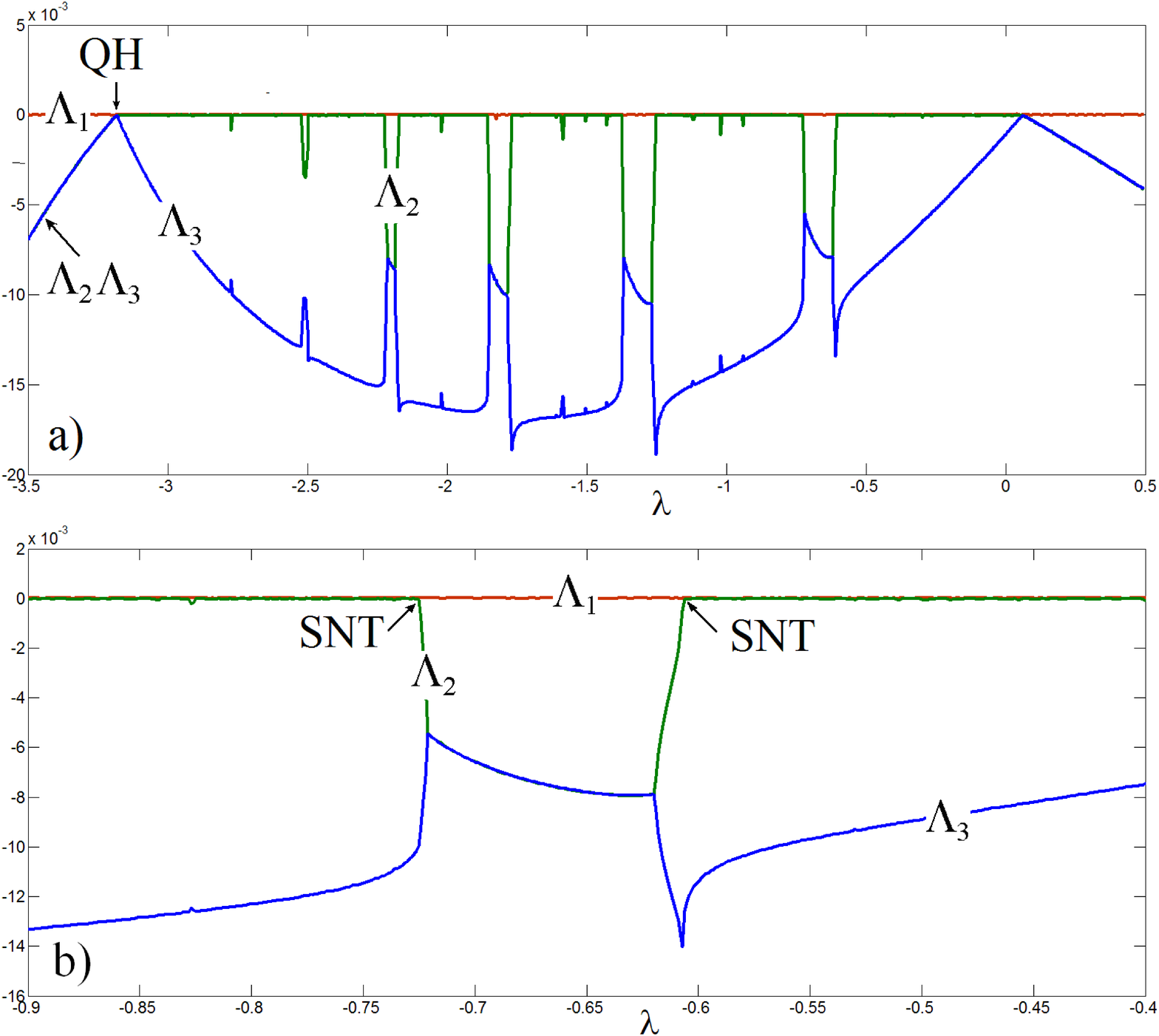}}
\caption{Three Lyapunov exponents of the model (\ref{eq3}) versus parameter $\lambda $
(a) and the magnified fragment (b). It is showed the location of
quasi-periodic Hopf bifurcation point QH and saddle-node torus bifurcations
SNT. Discretization parameter $h = 0.05$, parameter $\beta = 0.057$.}
\end{figure}

Let us now discuss ifurcations of quasi-periodic regimes. With this goal we
turn to Lyapunov exponent plots in Fig.3 calculated along selected in Fig.2
line $\beta = 0.057$. Such line crosses three-frequency periodicity region
from the bottom to the top. At the point $QH$ a two-frequency torus
($\Lambda { }_1 = 0)$ undergoes bifurcation. As can be seen from Fig.3, the
feature of this bifurcation is that below the threshold the exponents $\Lambda _2 $
and $\Lambda _3 $ are equal, $\Lambda { }_2 = \Lambda _3 $. At the
bifurcation point both these exponents vanish. Beyond the bifurcation point
exponents do not coincide: the second exponent is equal to zero, $\Lambda {
}_2 = 0$, and the third one becomes negative, $\Lambda _3 < 0$. Exactly at
the point of bifurcation the condition $\Lambda _1 = \,\Lambda { }_2 = 0$ is
fulfilled and a three-frequency torus emerges. This bifurcation is called
quasi-periodic Hopf bifurcation $QH$. Its distinguishing feature is the
realization of criterion of coincidence of two exponents beyond the
bifurcation point \cite{Bro08,Vit11}.

With increasing of control parameter $\lambda $ the route $\beta =
const\mbox{ }$on the chart of Fig.1 crosses many tongues of two-frequency
tori. On the plot of Fig.3a such tongues manifest themselves as dips in the graphs of the second exponent. The boundaries of these areas are formed by lines of
saddle-node torus bifurcation SNT. One of the deepenings in the enlarged view
is demonstrated in Fig.3b. Distinct characteristic of SNT bifurcation is
that the second Lyapunov exponent $\Lambda { }_2$ vanishes, but values
$\Lambda { }_2,\Lambda _3 $ are not equal to each other \cite{Bro08,Vit11}. Herewith
the third exponent $\Lambda _3 $ remains always negative (see Fig.3b). On the other side of the tongue, such a bifurcation takes place in
reverse order.

Accordingly we can indicate the curve of quasi periodic Hopf bifurcation QH in
Fig.2 separating three-frequency and two-frequency regions. In the
discrete model (\ref{eq3}) such a line replaces the line of Neimark-Sacker bifurcation
NS in the flow-prototype (\ref{eq1}) displayed in Fig.1.

\begin{figure}[!ht]
\centerline{
\includegraphics[height=16cm, keepaspectratio]{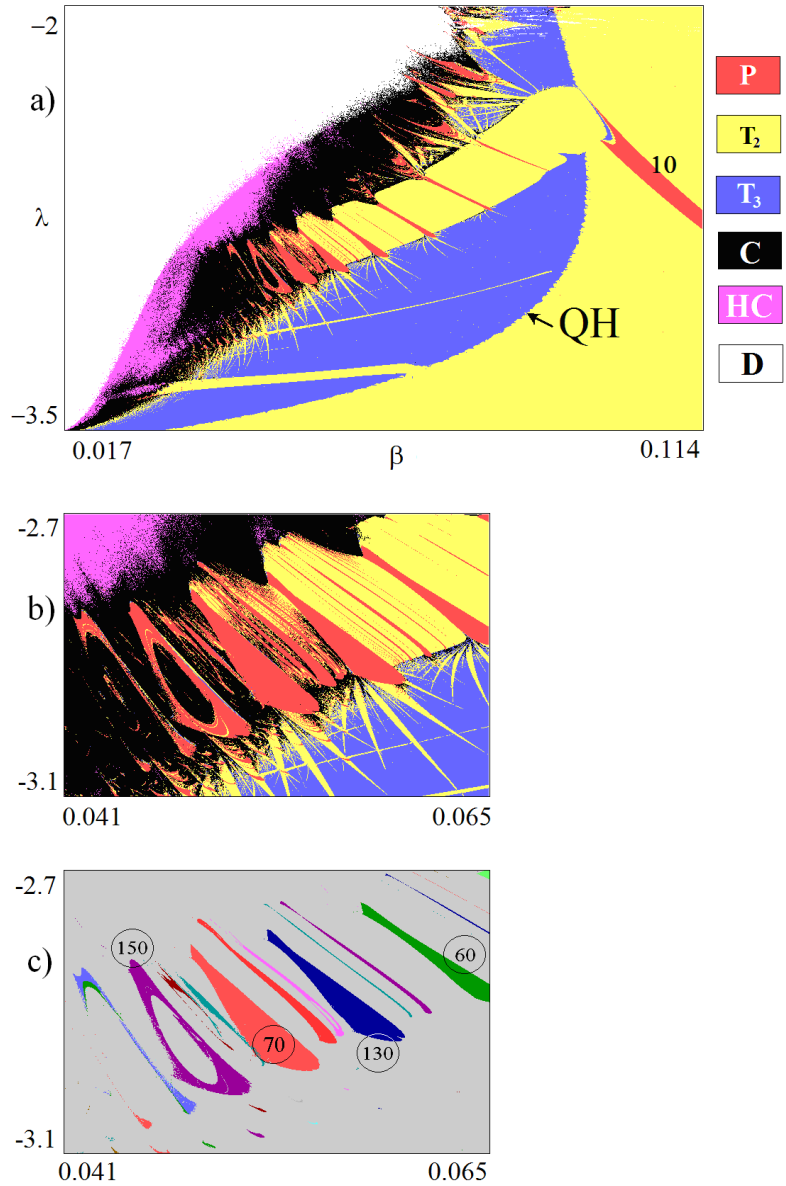}}
\caption{Lyapunov chart of torus map (\ref{eq3}) for parameter of discretization $h =
0.1$ (a), its enlarged segment (b), the chart of regimes of torus map (c).}
\end{figure}

It should be noted that quasi-periodic Hopf bifurcation is essentially
three-parameter phenomenon in contrast to the traditional two-parameter Neimark-Sacker
bifurcation. The physical nature of this fact is explained by adding of
parameter associated with additional frequency. In model (\ref{eq3}) such a
parameter can be a discretization step $h$, which is responsible for an
additional time scale. Therefore, we will increase the parameter $h$ and
trace the emerging changes in the structures of domains on the examined parameter plane.

Let us discuss the case $h = 0.1$, Fig.4. There are significant qualitative
changes on the parameter plane. Distinctive set of Arnold-type tongues is
disappeared. From the line of a quasi-periodic Hopf bifurcation QH the bands
of two-frequency regimes are issued. The crosswise bands of periodic
regions (exact resonances) are built in these two-frequency areas. On the
enlarged part of Lyapunov chart, Fig.4b, we see that mentioned resonances
generate secondary sets of the fan-shaped two-frequency tongues, immersed in
a three-frequency region.

To better visualize and distinguish periodic regimes we mark by different
colors and numbers the cycles of various periods (Fig.4c). Gray color
corresponds to all the non-periodic regimes. It can be seen that the
built-in areas of periodic regimes has different periods, at that their
value is big enough.

One more representative fact in Fig.4 consists in approaching of period-10
tongue to discussed two-frequency band from the main two-frequency area as
described in \cite{Bro08a}.

In Fig.5 we demonstrate the examples of phase portraits. There the
cycle of period 10 can be seen and its transformations within corresponding
two-frequency resonance region. An emergence of small isolated
ovals may be observed around the elements of the period-10 cycle. Cycles of very high
periods inside narrow regions of periodic regimes are very typical. Thus,
the structure of parameter plane is different from Fig.2.

\begin{figure}[!ht]
\centerline{
\includegraphics[height=10cm, keepaspectratio]{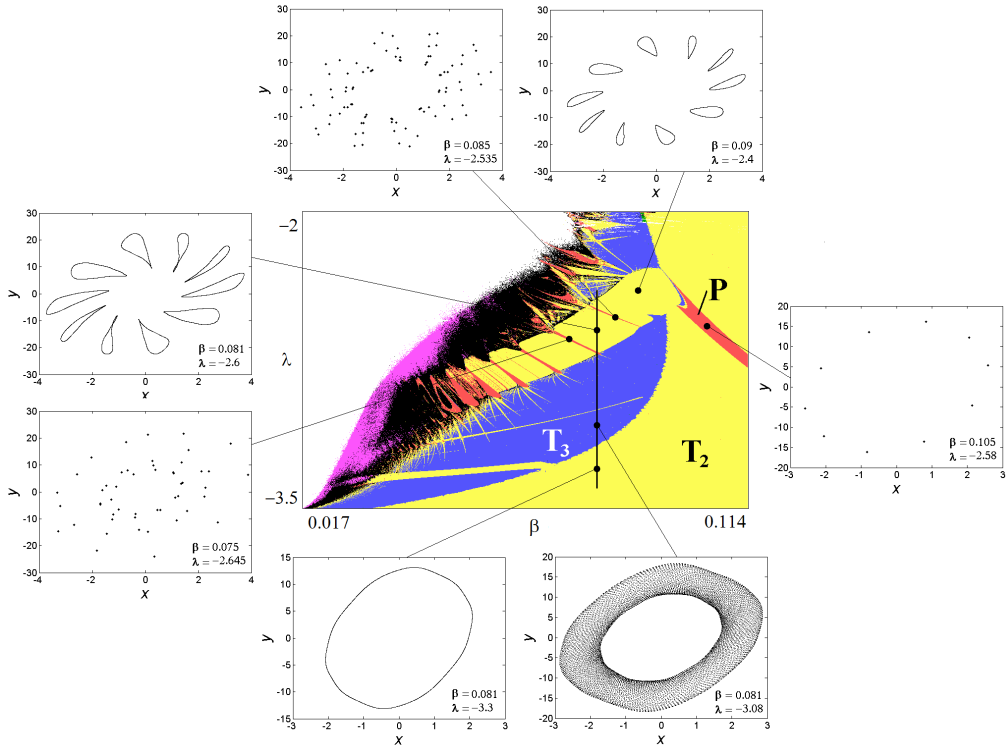}}
\caption{Attractors of torus map (\ref{eq3}). Parameter of discretization $h = 0.1$.}
\end{figure}

Plots of Lyapunov exponents calculated along vertical line in Fig.5 are
shown in Figure 6. In this case we observe a quasi-periodic Hopf bifurcation
QH. Another illustration of such bifurcation is a bifurcation tree presented
in Fig.6b in appropriate scale. One can note "smeary" crown of the tree that
signalized about quasiperiodic dynamics. At the point of quasi-periodic Hopf
bifurcation QH we observe the widening of bifurcation tree, and it occurs in soft way.

\begin{figure}[!ht]
\centerline{
\includegraphics[height=11cm, keepaspectratio]{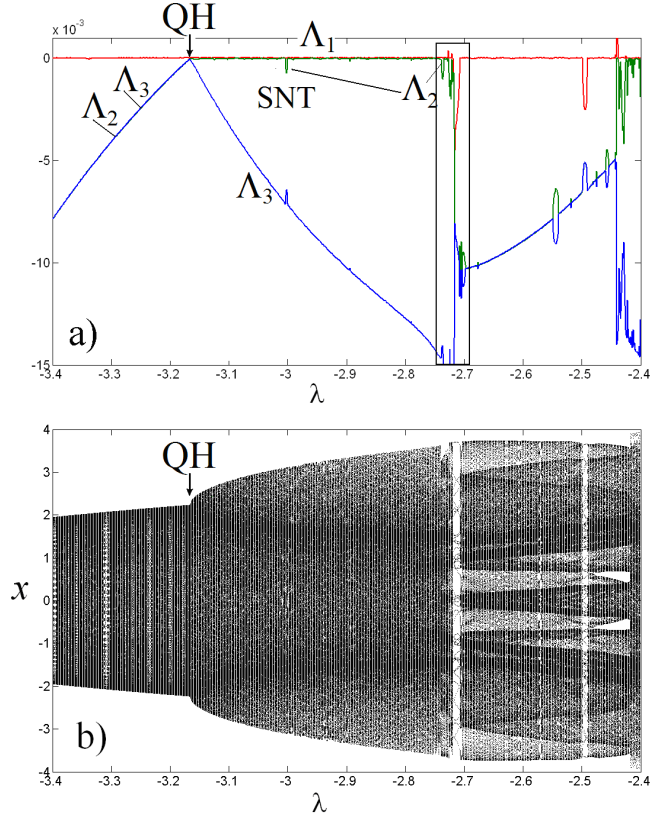}}
\caption{Dependence of three Lyapunov exponents on parameter $\lambda $ (à)
and bifurcation tree (b), numerically calculated for the model (\ref{eq3}). The
point of quasi-periodic Hopf bifurcation QH is marked. Discretization step $h
= 0.1$,parameter $\beta = 0.081$.}
\end{figure}

\begin{figure}[!ht]
\centerline{
\includegraphics[height=14cm, keepaspectratio]{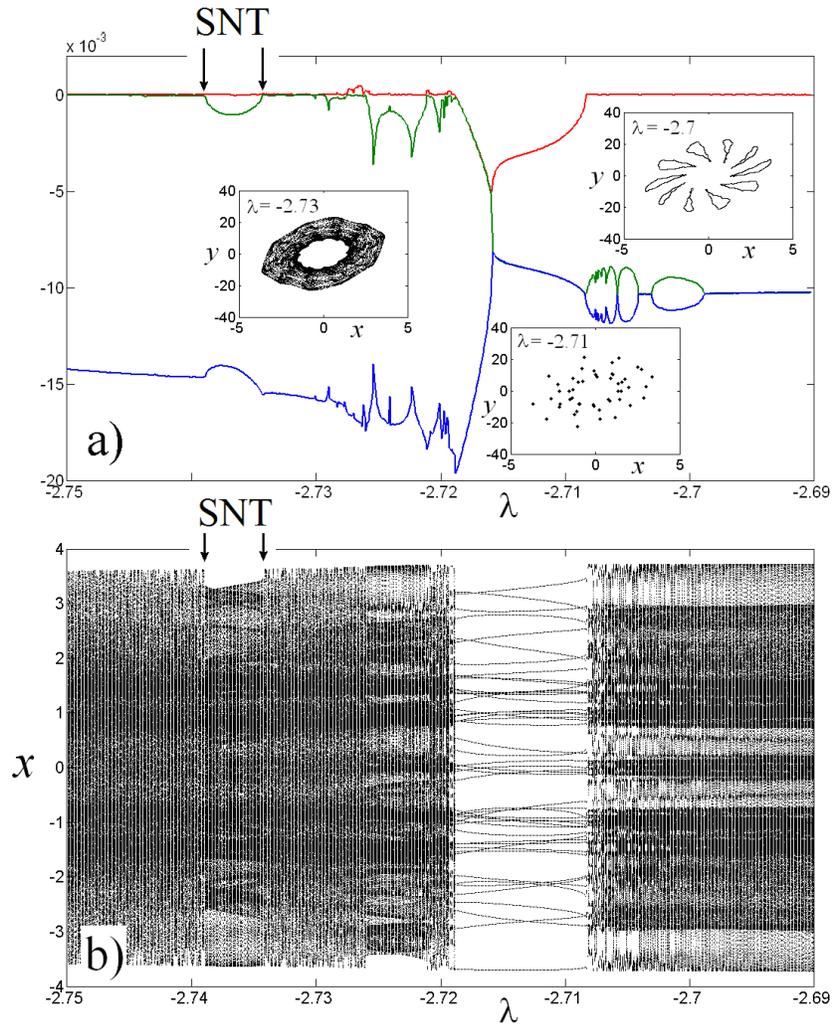}}
\caption{Enlarged fragment of Fig.6. Points of saddle-node torus bifurcation
SNT and typical phase portraits are indicated.}
\end{figure}

Figure 7 presents the enlarged fragments of Fig.6. We can see a strong
irregularity of diagrams due to the increasing complexity of parameter
plane. There are many alternating regions of two-frequency quasi-periodic and
periodic regimes. Nevertheless, there are distinguished areas of resonance
tori bounded by saddle-node bifurcation points SNT.

In Fig.7b we show the part of bifurcation tree to conclude that in contrast
to the points QH at saddle-node torus bifurcation SNT points the expansion
of the crown occurs abruptly. It happens due to the nature of the
bifurcation - stable and saddle two-frequency tori collide and
three-frequency torus occurs abruptly \cite{Bro08, Vit11}. On the bifurcation tree it is
also clearly seen resonance window of periodic regime.

Chart of Lyapunov exponents for greater value of parameter $h$ is shown in
Fig.8. Now the band of period 10 is not visible. The bottom of the largest
two-frequency tongue is limited by the bifurcation line, at that the
corresponding Lyapunov exponent vanishes. On the chart this line appears as a
thin blue strip. Its type can be determined via studying the phase portraits:
this is a torus-doubling bifurcation \cite{Ani95,Vit11}. Indeed, instead of a single
oval we have two overlapping ovals. They are obtained by projection of two
isolated closed invariant curves laying in three-dimensional space $\left(
{x,y,z} \right)$ on the surface of torus. The phase portrait in Fig.8
illustrates another such bifurcation for already doubled torus. Thus, at
increasing of the discretization parameter a new transformation occurs in
comparison with Fig.4.

\begin{figure}[!ht]
\centerline{
\includegraphics[height=9cm, keepaspectratio]{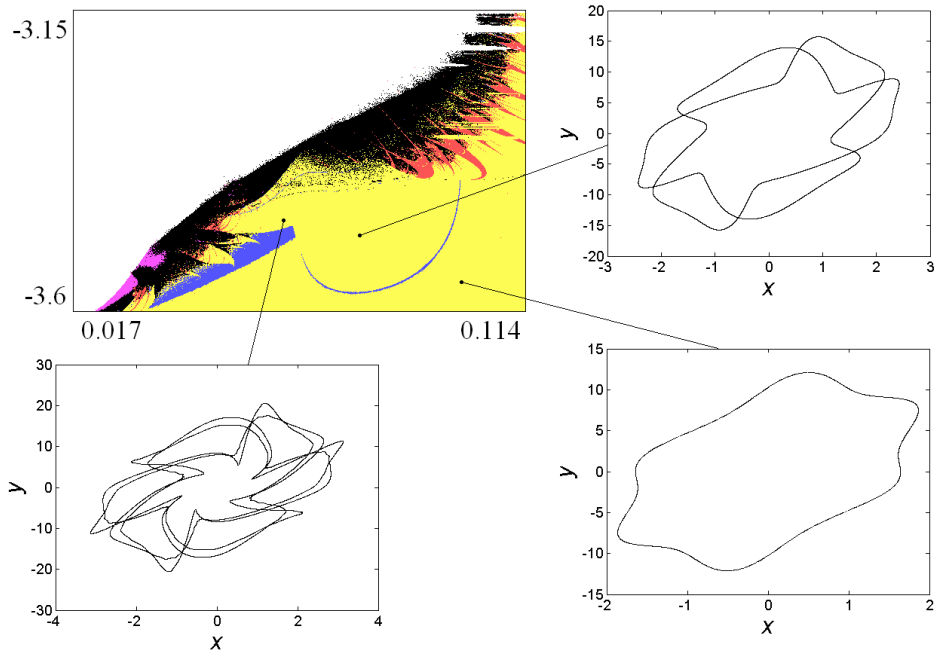}}
\caption{Chart of Lyapunov exponents and phase portraits of torus map (\ref{eq3}) at
$h = 0.16$.}
\end{figure}

\section{Non-autonomous model with quasi-periodic dynamics}

It is interesting to compare the studied dynamics with the case of
non-autonomous systems. It is necessary to choose such a
system to investigate that demonstrates two-frequency quasi-periodicity and Neimark-Sacker bifurcation.
An appropriate example is a universal map \cite{Kuz12} for which there are discovered all main
bifurcation scenarios of two-dimensional maps:
\begin{equation}
\label{eq4}
\begin{array}{l}
 x_{n + 1} = Sx_n - y_n - (x_n^2 + y_n^2 ), \\
 y_{n + 1} = J + x_n - \frac{1}{5}(x_n^2 + y_n^2 ). \\
 \end{array}
\end{equation}

In Fig.9 we demonstrate a diagram for the universal map \cite{Kuz12}, which combines
the properties of Lyapunov exponent chart and chart of periodic regimes. It
contains Neimark-Sacker bifurcation line NS, $J = 1$, and a set of Arnold
regular tongues.

\begin{figure}[!ht]
\centerline{
\includegraphics[height=7cm, keepaspectratio]{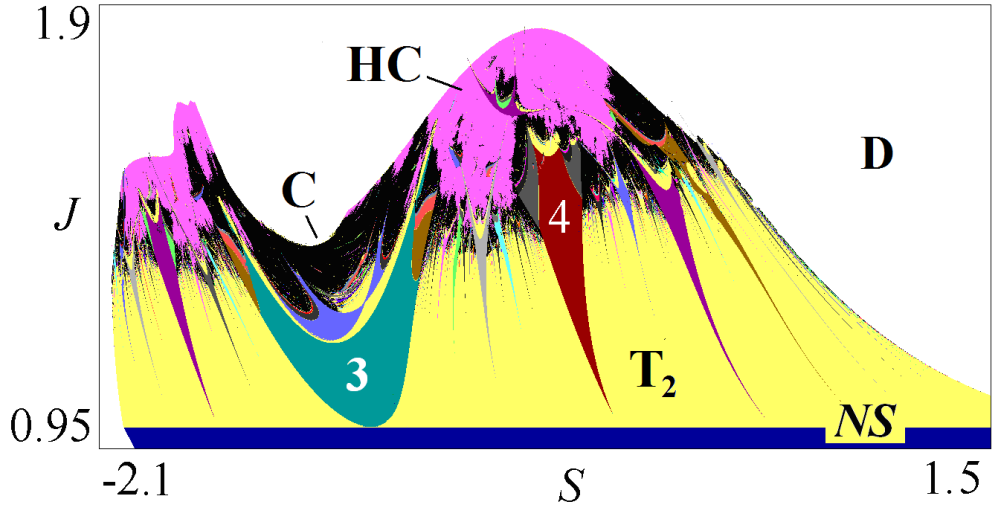}}
\caption{Combined chart of Lyapunov exponents and chart of periodic regimes
for universal two-dimensional map (\ref{eq4}), NS -- line of Neimark-Sacker
bifurcation.}
\end{figure}

Let us use the next form of external driving:
\begin{equation}
\label{eq5}
\begin{array}{l}
 x_{n + 1} = Sx_n - y_n - (x_n^2 + y_n^2 ), \\
 y_{n + 1} = J + x_n - \frac{1}{5}(x_n^2 + y_n^2 ) + \varepsilon \cos 2\pi
\theta _n , \\
 \theta _{n + 1} = w + \theta _n \mbox{ (mod1).} \\
 \end{array}
\end{equation}

A parameter determining the frequency of driving force is equal to the golden
mean, $w = \frac{\sqrt 5 - 1}{2}$, that provide the quasi-periodic type of
observed dynamics.

Lyapunov chart of model (\ref{eq5}) is shown in Fig.10. Compared with Fig.9 the periodic
regimes are replaced by two-frequency regimes, and two-frequency regimes
becomes three-frequency ones in Fig.10. Also we can see bands of
two-frequency modes replacing the tongues as well as typical ovals in bottoms
of some of the wide tongues corresponding to torus-doubling bifurcations. However,
a significant difference from the dynamics discussed above is the lack of
periodic resonances within tongues. Accordingly, there are no secondary
two-frequency resonances in their neighborhood.

Thus, the non-autonomous systems exhibit some characteristics that are
typical for autonomous models with quasi-periodicity, but fundamental
differences are inevitable.
\begin{figure}[!ht]
\centerline{
\includegraphics[height=7cm, keepaspectratio]{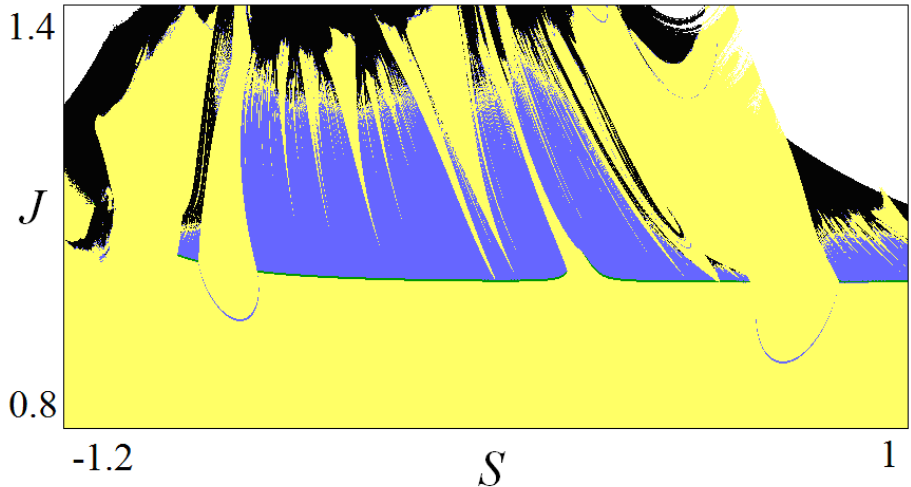}}
\caption{Lyapunov chart of universal two-dimensional map (\ref{eq5}) with
quasi-periodic external force. Parameter $\varepsilon = 0.1$.}
\end{figure}

\section{Conclusion}

Thus, the substitution of time derivatives by finite differences in the equation of generator of quasi-periodic oscillations (\ref{eq1}) provides a new
convenient model in the form of three-dimensional map. This map demonstrates
the regimes of two-frequency and three-frequency quasi-periodicity and all
basic quasi-periodic bifurcations. Typical Lyapunov exponent plots and
bifurcation trees are presented. Such map allows to study three-parameter
structure of quasi-periodic Hopf bifurcation. With variation of third
parameter there are significant changes in the structure of parameter plane
and the form of phase portraits. The fundamental difference from non-autonomous systems with quasi-periodic forcing is the possibility of various
periodic resonances and associated with them secondary tongues of
two-frequency regimes.

\textit{The research was supported by Russian Foundation for Basic Research RFBR (grant No 15-02-02893).}

\end{document}